\documentstyle[preprint]{aastex}
\slugcomment{Astronomical Journal}

\shorttitle{The Solar Neighbourhood XI}
\shortauthors{Deacon et al.}
\begin{document}
   \title{The Solar Neighbourhood XI: The trigonometric parallax of SCR 1845--6357}

   \author{Niall R. Deacon, Nigel C. Hambly}
\affil{Institute for Astronomy, University of
              Edinburgh, Blackford Hill, Edinburgh EH9 3HJ}
    \author{Todd J. Henry, John P. Subasavage, Misty A. Brown, Wei-Chun Jao}
   \affil{Department of Physics and Astronomy,Georgia State
          University,Atlanta, GA, 30303-3088}

   \begin{abstract}
We present a trigonometric parallax for the nearby star
SCR1845--6357, an extremely red high proper motion object discovered by 
\citet{h6} and identified via accurate photoelectric photometry
and spectroscopy to be 
an M8.5 dwarf with a photometric parallax of $4.6\pm0.8$~pc by
\citet{h8}. Using methods similar to those described in \citet{d2} we have derived a full astrometric solution from 
SuperCOSMOS scans of eight survey and non--survey 
Schmidt photographs held in the United Kingdom
Schmidt Telescope Unit plate library. We calculate the trigonometric
parallax to be $\pi=282\pm23$~mas yielding a distance of
$3.5\pm0.3$~pc which implies an absolute K$_{s}$ magnitude of 10.79. This distance calculation places SCR1845--6357
as the 16th closest stellar system to the Sun.
   \end{abstract}

\section{Introduction}
There has been a recent flurry of discoveries of stars within the
 local solar neighbourhood. Some have been discovered as companions to
 well--known
 nearby stars (eg.\ $\epsilon$ Indi B/C; see \citealp{m2} and
 references therein) while others are
 entirely new systems (eg.\ \citealp{t2}).
The RECONS group (Research Consortium on Nearby Stars) has sought to
 identify and study all systems within 10pc of the Sun. 
 Previous papers of this series have used a variety of photometric and
spectroscopic techniques to identify and study samples of
 nearby stars and describe discovery of companions to known nearby stars
 and identification of entirely new systems.  
 Most recently, a new initiative \citep{h6} 
identified five nearby high proper motion stars, 
two of which have initial photometric
 parallaxes within the RECONS horizon. The nearest of these, SCR1845--6357,
(hereafter the target) was shown subsequently to have 
a photometric parallax of $4.6\pm0.8$~pc \citep{h8}. Photometric data for the target are reproduced in
 Table~\ref{phot}. These show that the object is very red; \citet{h8} classified the target as an M8.5 dwarf.

\section{Observational data and reduction}

Luckily there is a wealth of astrometric data
 available for this object, partly because it lies in the overlap region
between the standard Schmidt photographic survey fields, and partly because non--survey
programmes have fortuitously observed the field containing the target
frequently and over a long time baseline. The target 
 was identified on two United Kingdom Schmidt Telescope (UKST) R survey plates and one European Southern 
 Observatory (ESO) survey R plate. These had already been scanned as part of 
 the SuperCOSMOS Sky Survey (\citet{h2} and references therein). 
In addition six UKST non-survey R photographs from the UKST Plate
 Library were selected for their image quality and useful parallax factors, and were scanned especially for 
this study using SuperCOSMOS. Hence a total of nine Schmidt photographs 
were used, details of these are shown in 
 Table~\ref{plates}. These also provide a nine year baseline, ideal for
measuring an accurate proper motion.\\

   \begin{table}
      \caption[]{Photometric data on SCR1845--6357. JHK
      magnitudes are taken from the Two Micron All Sky Survey, V, R
      and I measurements are from Henry et al. 2004. Note that
 the B$_{\rm J}$ measurement comes from a plate where the target image is
 deblended and hence may not be as accurate as the other measurements.}
         \label{phot}
\begin{tabular}{ccccccccl}
\hline
B$_{\rm J}$&R$_{\rm 59}$&I$_{\rm N}$&J&H&K$_s$&V&R$_{C}$&I$_{C}$\\
\hline
19.05&16.33&12.53&9.54&8.97&8.51&17.40&15.00&12.47\\
\hline
\end{tabular}
   \end{table}
Each plate was scanned individually on SuperCOSMOS and processed using
standard methods (\citet{h2} and references therein). Global astrometric
plate solutions result in systematic errors of order 
$\sim0.2$~arcsec in absolute
 positions. Hence it is necessary to correct for these; a local linear
plate model with respect to the array of mean positions from all
measures was employed for this purpose. A sample of circular, single, stellar images which are not 
affected by proximity to bright stars were selected as local astrometric
reference stars. These reference stars were  used to fit linear models
for each of the plates with respect to the array of mean reference
star positions. The residual errors
from the reference stars after these models are applied give an
indication as to the astrometric quality of each plate; as a result of
this test, one of the non--survey plates was excluded from the rest of
the study owing to poor astrometric quality (see Table~\ref{plates}).\\ 

It is also important that the  maximum distance of reference stars from the target is carefully chosen. 
Too small a distance and the number of reference stars will be too
few, too large and non--linear plate effects may leave systematic
errors in the linear fit (indicated by an increase the RMS error). In order to choose the maximum
reference star distance correctly several different selections were made 
and the average RMS errors (the mean of all the errors from all the plates) from the local linear plate fit were noted along with the 
 number of reference stars. The results are shown in Table~\ref{Ref}. We chose the maximum extent available (20~arcminutes) from the
area in common between both plates as there is no indication of an
increase in systematic errors up to this value.
   \begin{table*}
      \caption[]{Schmidt photographs used in this study; relative
   astrometric quality is indicated by the $\sigma_x$ and $\sigma_y$ values(see text). One plate was excluded due to poor astrometric quality.}
         \label{plates}
\begin{tabular}{cccccccccl}
\hline
Plate & Date     & LST &Zenith& Emul- & Filter & Exp.  & $\sigma_x$ & $\sigma_y$ &Material\\
No.   & (ddmmyy) &     &Angle& sion  &        & (min) & (mas)      & (mas)      &\\
\hline
\multicolumn{10}{c}{Plates Used}\\
\hline
ESOR6887&30/04/87&1713&$37.7^{\circ}$&IIIaF&RG630&120&46&64&Glass copy
of \\
&&&&&&&&&ESO survey plate\\
OR13751&20/06/90&1843&$32.7^{\circ}$&IIIaF&OG590&60&41&42&Original
non--survey\\
&&&&&&&&& plate\\
OR14370&16/06/91&1831&$32.8^{\circ}$&IIIaF&OG590&55&40&39&Original
non--survey\\
&&&&&&&&& plate\\
OR15689&12/08/93&1746&$34.0^{\circ}$&IIIaF&OG590&55&40&42&Original
survey \\
&&&&&&&&&plate\\
OR16753&28/08/95&1850&$32.7^{\circ}$&IIIaF&OG590&80&46&44&Original
survey \\
&&&&&&&&&plate\\
OR17012&24/03/96&1648&$37.5^{\circ}$&4415&OG590&15&48&49&Original
non--survey\\
&&&&&&&&& film\\
OR17038&15/04/96&1757&$33.6^{\circ}$&4415&OG590&15&53&56&Original
non--survey\\
&&&&&&&&& film\\
OR17256&15/09/96&1838&$32.7^{\circ}$&4415&OG590&15&55&64&Original
non--survey\\
&&&&&&&&& film\\
\hline
\multicolumn{10}{c}{Plate Not Used}\\
\hline
R 5991 &16/05/80&2032&$36.7^{\circ}$&IIIaF&RG630&150&107&92&Original non--survey\\
&&&&&&&&& plate\\
\hline
\end{tabular}
   \end{table*}
%

   \begin{table}
      \caption[]{Selecting the optimum maximum distance of reference
   stars from the target. Note the number of reference
      stars does not increase as the square of the maximum radius due
      to plate boundary cutoffs.}
         \label{Ref}
\begin{tabular}{ccl}
\hline
Maximum distance&Average RMS&Number of\\ 
from target&error&reference stars\\
(Arcminutes)&(mas)&\\
\hline
20&56&108\\
18&56&108\\
16&56&98\\
14&58&77\\
12&59&56\\
10&58&41\\
\hline
\end{tabular}
   \end{table}

\section{Results}
\begin{figure}
\includegraphics[scale=0.9]{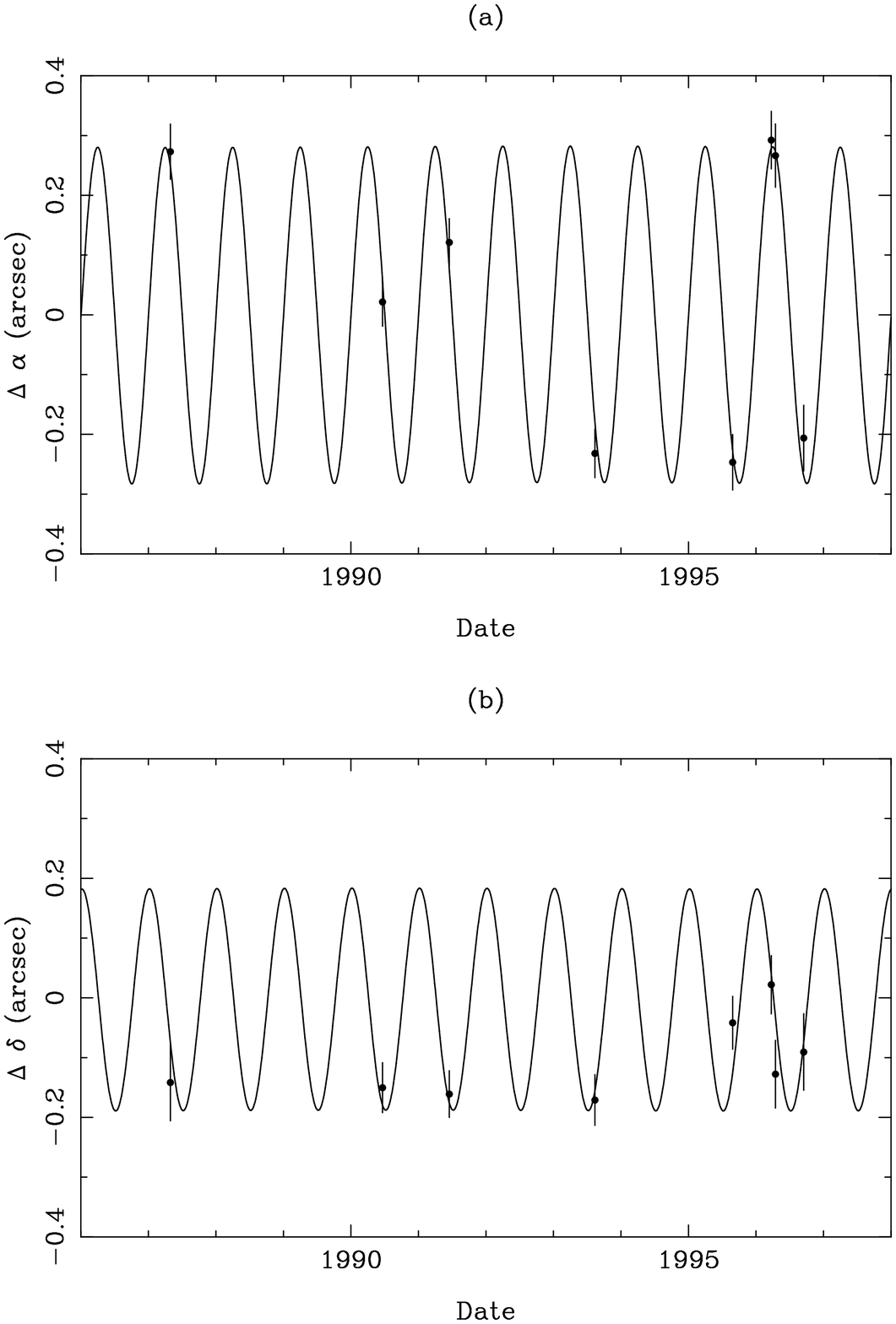}
      \caption{The deviation from the proper motion in (a) RA and (b) Dec 
      versus time. The line shown is the path of parallax motion predicted
         by the astrometric solution}
         \label{parallaxa}
   \end{figure}

\begin{figure}
\plotone{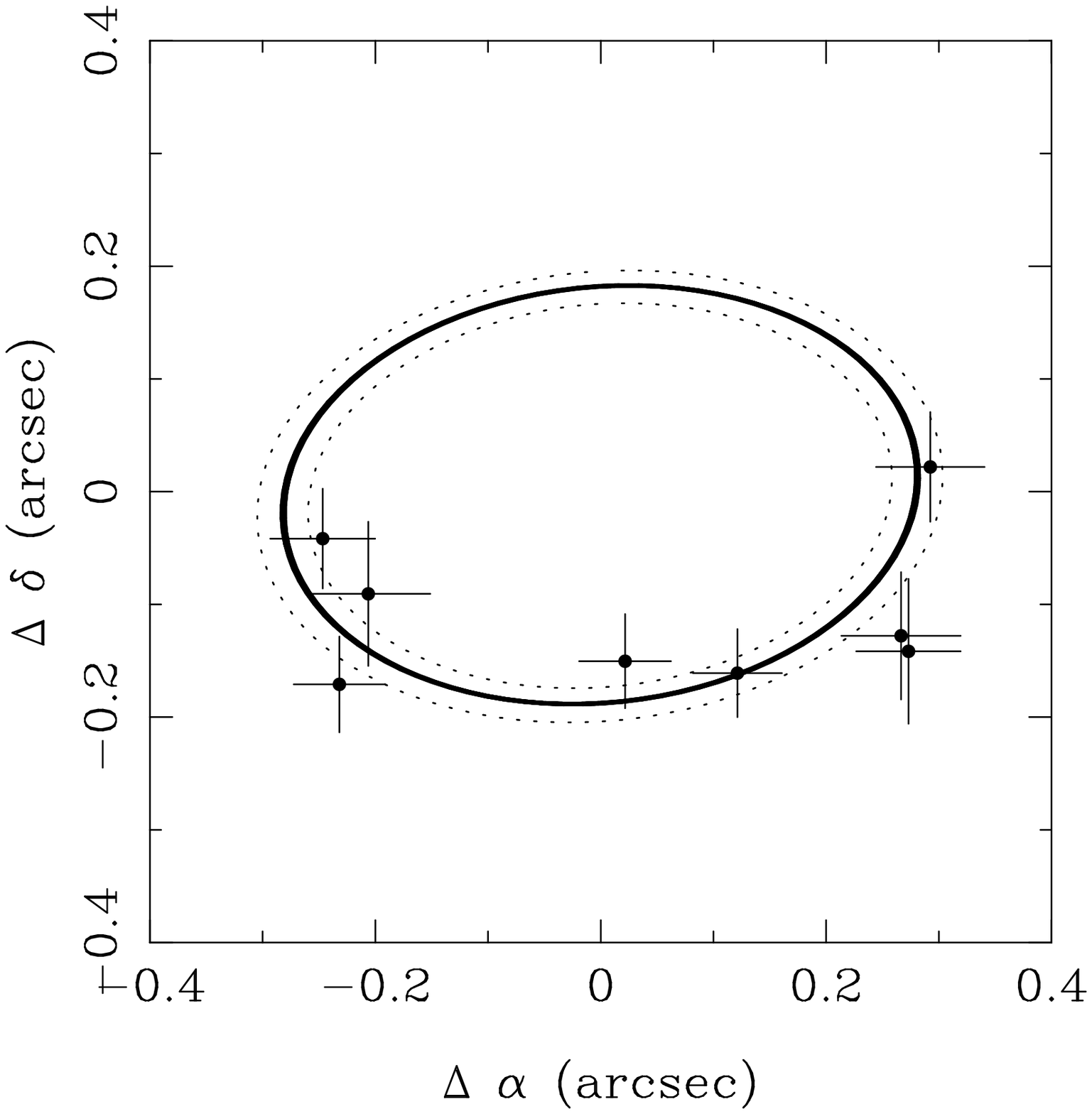}
      \caption{The parallax ellipse traced out by the target, the dotted
   lines represent one sigma upper and lower limits on the parallax.}
         \label{parallaxb}
   \end{figure}

Once the process described in Section~2 was carried out,
 plate--to--plate systematic errors were eliminated, the relative astrometric
quality of each plate with respect to the mean positional measures was
determined and it was possible to apply a 
 full weighted astrometric fit to the target and reference stars.
 Equations~\ref{RAE} and~\ref{DecE} show the astrometric models
 for Right Ascension and 
 Declination respectively (where $f_{\alpha}$ and $f_{\delta}$ are the
 parallax factors in Right Ascension and Declination):
\begin{equation}	
\alpha = \alpha_{o} + \mu_{\alpha}t + \pi f_{\alpha};
\label{RAE}
\end{equation}
\begin{equation}	
\delta = \delta_{o} + \mu_{\delta}t + \pi f_{\delta}.
\label{DecE}
\end{equation}
The error--weighted design matrix formed from these equations was solved via
Singular Value Decomposition to yield a least--squares fit; we
employed numerical routines from the SLALIB positional astronomy 
library \citep{w2}. 
The results for the target
(along with $\chi_{\nu}^{2}$, ie.\ $\chi^{2}$ normalised per
 degree of freedom) are shown 
 in Table~\ref{Res}. The total proper motion is $2.64 \pm 0.0082$ arcseconds/year with a position angle
of 74.9 degrees, these compare with the  values
of $2.56 \pm 0.013$ arcseconds/year and 74.8 degrees quoted in
\citet{h6}. The proper motion measurements differ but this may be due
to systematic errors in either measurement or because \citet{h6} did
not include a parallax in their astrometric solution. The
trigonometric parallax was found to be $\pi = 282  \pm 23$ milliarcseconds. Figure~\ref{parallaxa} shows
 the deviation in position of the target from it's proper
 motion. Figure~\ref{parallaxb} shows the parallax ellipse traced out
 by the target star.   

   \begin{table}
      \caption[]{The full astrometric solution for SCR1845--6357, note
      the $\chi_{\nu}^{2}$ indicating a good fit to the model.}
         \label{Res}
\begin{tabular}{cccl}
\hline
Parameter&Fitted Value&Error&Units\\
\hline
RA on 01/01/2000 & $18^{\rm h}45^{\rm m}05^{\rm s}\!.2008$ & 0.0407~as& --- \\
Dec on 01/01/2000& $-63^{\circ}57'47"\!.355$ & 0.0435~as& --- \\
$\mu_{\alpha}\cos\delta$&2.5495&0.0055&as/yr\\
$\mu_{\delta}$&0.6874&0.0061&as/yr\\
$\pi$&282&23&mas\\
$\chi_{\nu}^{2}$&0.49& --- & --- \\
\hline
\end{tabular}
   \end{table}

\begin{figure}
\plotone{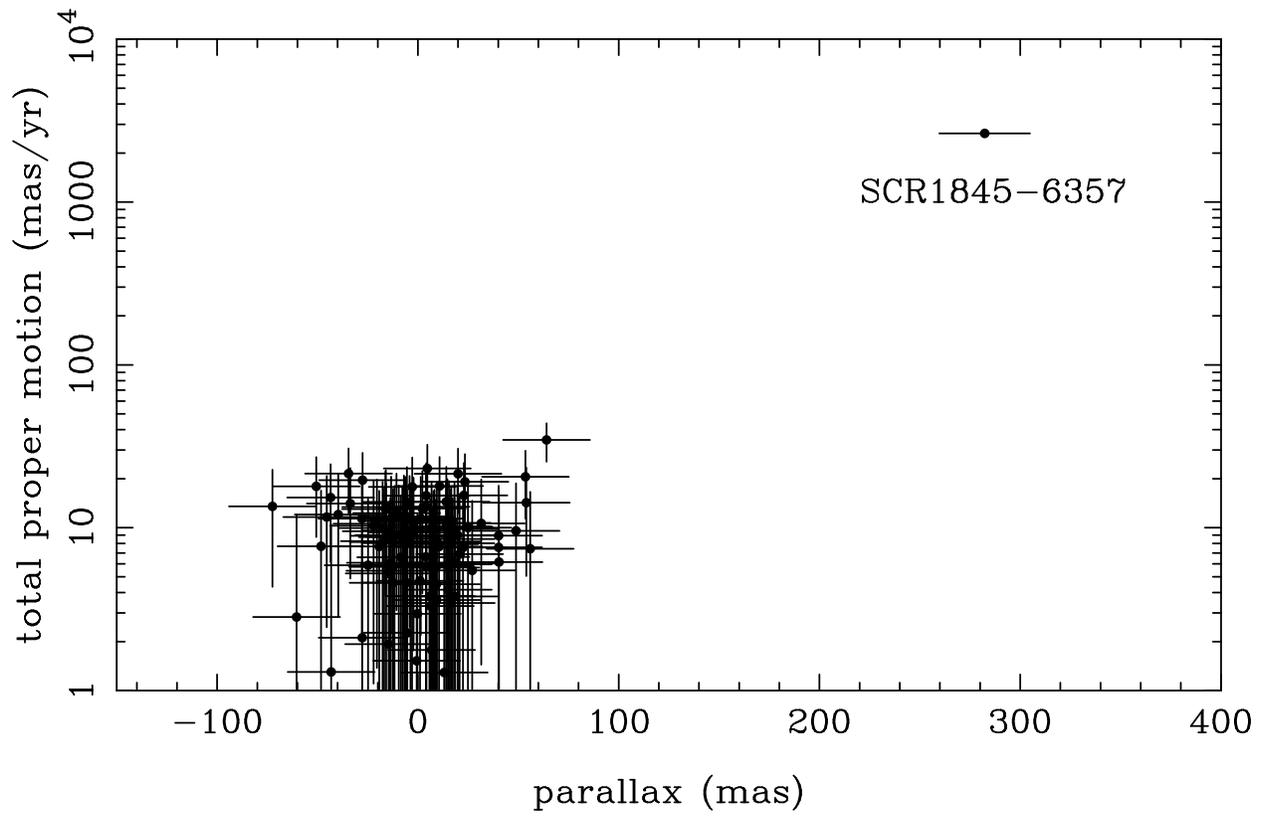}
      \caption{Comparing the fitted parallax and proper motions for
   the reference stars with SCR1845--6357.}
         \label{paraplot}

   \end{figure}

As a test of the astrometry, each reference star was also run through the
same astrometric model fitting procedure and the proper motions and
parallaxes of the reference stars were found. 
 These are plotted in Figure~\ref{paraplot} along with the target. It is clear that the target is well separated from the mass of reference
 stars. In general the reference stars have
proper motions and
parallaxes consistent with zero (and a mean value of $\chi_{\nu}^{2}\sim$0.72
indicating good model fits) and hence are good reference stars. To further
investigate the reference star distances, and for the purposes of
correcting the measured (ie.\ relative) parallax to an absolute parallax, 
the mean B$_{\rm J}$--R$_{59}$
and~R$_{59}$ for the reference stars were found. The mean B$_{\rm J}$--R$_{59}$
 calculation was then used to
 find that the mean spectral type (assuming they are dwarfs) for the
reference stars is G5~\citep{z1}. From this
 the mean expected absolute R$_{59}$ magnitude~\citep{z1} can be deduced which combined
with the mean R$_{59}$ magnitude yields the mean expected distance of the reference stars. The mean expected trigonometric parallax of the reference stars was
 thus found to be $\pi=0.7$~mas. This is clearly insignificant compared to the
 formal error on our parallax for the target, so we made no correction
 from relative to absolute parallax.

To further
 test that the method used was sound, 100 sets of simulated
 data were created. Each had the same astrometric parameters as the
 target and each data point was given a random Gaussian
 error calculated from the RMS error estimates of the particular
 plate. These were then run through the astrometric solution fitting
 program. No significant offsets were found and the error on the
 parallax calculated from the scatter of the simulated data solutions
 was in good agreement to that predicted by the astrometric solution. 

\begin{figure}
\plotone{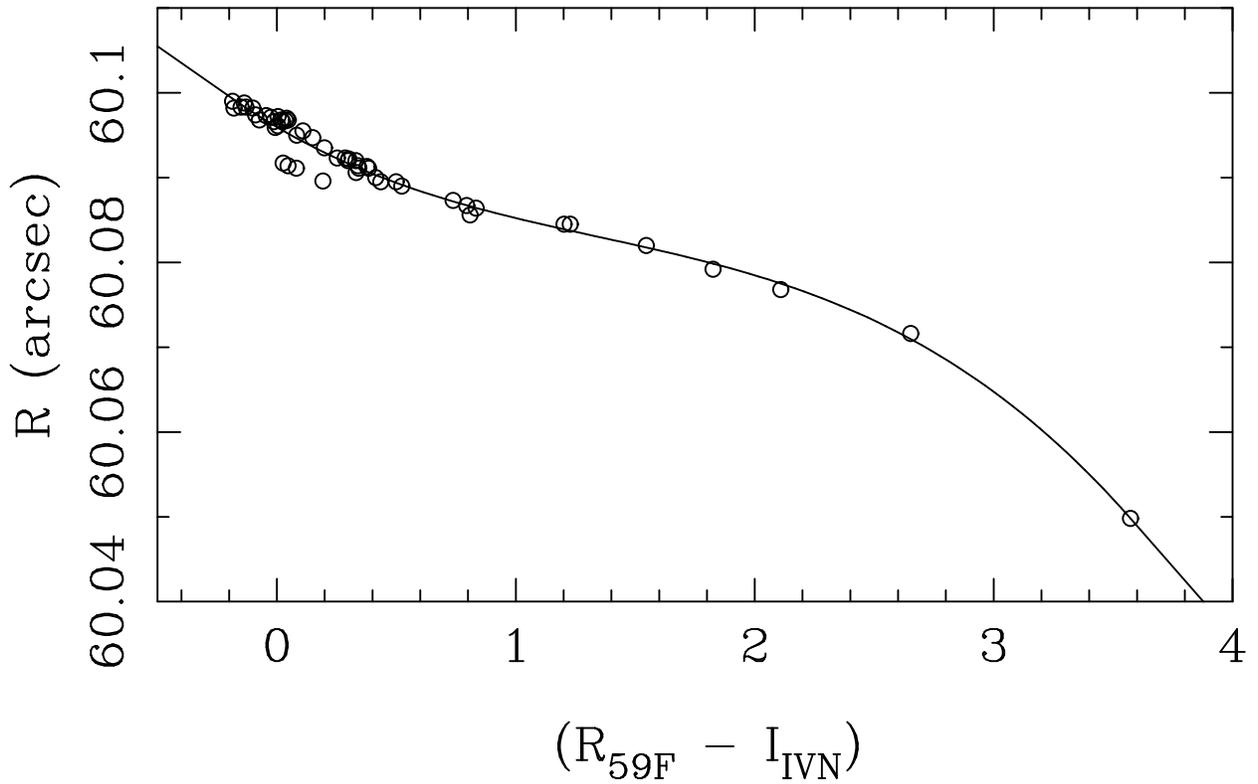}
      \caption{Computational models (after Hambly et al.~2001b and
references therein) for the coefficient of refraction $R$ at an
airmass of 1.5  as a function
of synthetic photographic (R--I) colour. Data are from a spectrophotometric
atlas except for the point at (R--I, $R$)=(3.6,60.05) where we have used
the spectrum for SCR1845--6357 presented in Henry et al.~(2004). The solid
line is a polynomial fit to the data points.}
         \label{dcr1845}
   \end{figure}

Finally, we examined differential colour refraction (DCR) effects between
the reference stars and the target. In Figure~\ref{dcr1845} we show model
predictions for the coefficient of refraction, R, for stars of various
synthetic photographic colour. We used methods described in \citet{h4}
and references therein; we additionally used the flux calibrated
spectrum of SCR1845--6357 presented in Henry et al.~(2004) to compute~R$_{59}$
and synthetic (R$_{59}$--I$_{N}$) for the extremely red target. The reference stars are taken as an ensemble of points that
    typically have (R$_{59}$--I$_{N}$)~$\sim0.5$, whereas the target has (R$_{59}$--I$_{N}$)=3.6
computed from the spectrum, and 3.75 from the photographic photometry
listed in Table~\ref{phot} (these colours are consistent within the 
photographic photometric errors of $\sim0.1^{\rm m}$ for R$_{59}$--I$_{N}$). These
colours indicate a difference in refraction coefficient of 
$\Delta$ R $\sim35$~mas at airmass of 
1.5 (Figure~\ref{dcr1845}). This difference changes negligibly with airmass,
    i.e. recomputations at 1.0 and 2.0 airmasses show a zero point
    shift in the locus of Figure~\ref{dcr1845} , but no significant change in the
    Delta R value for the two colors. 
However the quantity of most relevance is of course how this differential
refraction coefficient translates into image displacement at the different
hour angles and zenith distances of the observations in Table~\ref{plates}.
Typically, the DCR shift for RA is $\pm5$~mas and for Dec is $\pm1.5$~mas;
the largest change between any two observations is 10~mas for RA and 3~mas
for Dec because all of the plates used are within 2 hours of
    the meridian, so DCR is minimal. Hence, we feel justified in neglecting these corrections in our
astrometric model since the positional uncertainties dominate systematic
effects due to DCR. 

\section{Discussion}
The calculated trigonomteric parallax for the target gives a distance
of $3.5\pm0.3$~pc. Consulting the RECONS list of nearby
stars\footnote{http://www.chara.gsu.edu/RECONS/}
we find that this makes SCR1845--6357
the 16th nearest stellar system to the Sun. The upper and lower one
sigma error boundaries would make it the 23rd and 10th nearest stellar
system respectively. We note that the RECONS photometric parallax is
$4.6\pm0.8$~pc ~\citep{h8} hence our new trigonometric parallax
puts this object $2.8 \sigma$ nearer than was first estimated. \citet{h6} estimated the distance to be
$3.5 \pm 0.7$~pc; we expect the true distance of the target to be
between the estimate given here and that of \citet{h8}.

\section{Conclusion}
We have measured the trigonometric parallax of SCR1845--6357 and found
it to be the 16th nearest stellar system to the Sun. This demonstrates
that the wealth of astrometric information on the
     many plates taken over the past 50 years can yield new insight
     into the nearby star population and, in particular, permit the
     determination of valuable trigonometric parallaxes.
\section*{Acknowledgements}
The authors would like to thank Sue Tritton and Mike Read for their
help in selecting plates and Harvey MacGillivray and Eve Thomson for
their prompt scanning of the material on SuperCOSMOS. SuperCOSMOS is funded by a grant from the
UK Particle Physics and Astronomy Research Council.
This publication makes use of data products from the Two Micron All Sky Survey, which is a joint project of the
University of Massachusetts and the Infrared Processing and Analysis Center/California Institute of Technology,
funded by the National Aeronautics and Space Administration and the
National Science Foundation.

\label{lastpage}

\end{document}